\documentstyle[prl,aps,epsf]{revtex}
\begin{document}
\title{ Squeezed Light Generation in Nonlinear System with
Chaotic Dynamics}
\author{Kirill N. Alekseev}
\address{Theory of Nonlinear Processes Laboratory,
Kirensky Institute of Physics,
Russian Academy of Sciencies,
Krasnoyarsk 660036, Russia\\
E-mail: kna@iph.krasnoyarsk.su\\
E-mail: kna@vist.krascience.rssi.ru}
\maketitle
\begin{abstract}
The squeezing in a nonlinear system with chaotic dynamics
is considered. The model describing interaction of collection of
two-level atoms with a single-mode self-consistent field and
an external field is analyzed. It is shown that in semiclassical
limit, in contrast to the regular behaviour, the chaotic  dynamics
may  result  in:  (i) an  increase  in  squeezing,  (ii)  unstable
squeezing and contraction of time intervals of squeezing on large
enough times. The possibility of the experimental observation  of
the described effects is  discussed. The  obtained   results  are
rather  general  and  do  not  depend   on   the   model   under
consideration.
\end{abstract}
\section{Introduction}
     In recent years a great deal of interest has been focused on
the investigation of chaotic dynamics in different physical
processes [1,2]. Alongside with the study of dynamical chaos in
the classical systems, in the middle of 70s more and more
researchers began to realize the importance of the problem of
quantum chaos. The subject of quantum chaos is the study of some
peculiarities in the behaviour of quantum systems which are
classically chaotic [3-5]. A lot of interesting theoretical
results have been obtained in the investigation of quantum chaos.
One of the most important problem now is to study concrete
physical systems with quantum chaos. Such systems have to meet
certain requirements, i.e. they must be: a) semiclassical, b)
chaotic in the classical limit, c) Hamiltonian (nondissipative).
The last requirement is, apparently, not necessary but the study of
the dissipative quantum chaos has only begun [6].
\par
     A very promising models in the study of quantum chaos are
the models of quantum optics. First of all, these are models
describing interaction between light and a collection of atoms.
Really, the semiclassical conditions and those of (semi)classical
chaos can be easily met for such models (e.g., see review [7]).
But the majority of the optical systems with classical chaos are
dissipative (lasers, bistable devices, etc.). Up to now the only
exception was the BZT model [7,8] -- the semiclassical
Jaynes-Cummings model [9] without the assumption of the
rotating-wave approximation (RWA). But the global chaos arises in
the BZT model if only the RWA is violated \footnote[1]{\noindent
  The RWA is violated when the coupling constant between the
atoms and the field (in units of frequency) becomes of the order
of transition frequency.}
 ,and this requires
very large atom densities and values of dipole moments in a
experiments (numerical estimates see in [7,8]). Only recently
there appeared some modifications of Jaynes-Cummings model
assuming the Hamiltonian chaos in the semiclassical description
and within the framework of the RWA [10-13].
\par
     On the other hand, at present there are a number of purely
quantum (nonclassical) effects arising in the interaction of
light with atoms (squeezing, photon antibunching, etc.) [14-17]. As
a rule, these effects are more pronounced in the semiclassical
limit. However, all the nonclassical effects with light have
been studied for the integrable systems with regular behaviour
\footnote[2]{\noindent The problem of a squeezing in chaotic system was 
mentioned in
concluding remark of the paper [19]. But  no analysis of
the problem was given in [19].}.
Therefore, in order to investigate the problems of quantum
chaos, as well as those of quantum optics, it is of interest to
analyze the influence of the chaotic behaviour on the
nonclassical optical effects.
\par
     In this paper, a squeezed light generation in a nonlinear
system with chaotic behaviour is considered. The model [12] is
used as a concrete example, and the method of $1/N$ expansion into
modification suggested in [18] is applied.
\par
     The main result of the paper is the demonstration of the fact
that in the semiclassical limit chaos may increase squeezing of
light. A phenomenon of the unstable squeezing is described. The
unstable character of squeezing is manifested in a great change
of time intervals, during which the the squeezing is possible when
the initial conditions or parameters of the systems change
slightly.
\par
     The paper isorganized  as follows. Section 2
presents a brief description of the model [12] and the procedure
of $ 1/N$ expansion. Here also the equations are obtained which
describe light squeezing in the semiclassical limit. Section 3
deals with the conditions for transition to chaos in the model
discussed. Section 4 is devoted to the comparison of squeezing
for regular and chaotic dynamics. In Discussion other models of
quantum optics are considered permitting semiclassical chaos
and as consequence a possible enhance in squeezing. The
conditions for observation of chaotic squeezing in experiments are
also discussed.
\par
Preliminary results of the paper were presented in [20].
\section{ $1/N$ Expansion and Squeezing}
     Let a sample with a gas of $N$ two-level systems (TLS) be placed
in a ring single-mode, high-$Q$ cavity of resonant frequency $\omega$. Now
let us consider the interaction between $N$ two-level atoms, a
self-consistent field and an external, classical, amplitude-modulated
 field of the form
\begin{equation}
     E_{\rm ext}(t) = E_{0}  F(t)\cos \omega t
\end{equation}
injected into the cavity in the plane $z=0$. In eq. (1) $E_{0}$  is the
external field amplitude, and $F(t)$ is a function, slowly changing
in comparison with a carrier frequency $\omega$ and defining the
amplitude modulation. Using the approximation, known in the theory
of optical bistability and lasers with an injected external
signal as the ``mean field model'' [21], the Hamiltonian in the
interaction representation can be written in the form
\begin{eqnarray}
 H = N H_{N},& &  H_{N}=H_{JC}+H_{\rm ext},\\
 H_{JC} &=& i \hbar\omega_c (a^{+}J_{-}-aJ_{+}),\\
 H_{\rm ext} &=& i\hbar\omega_c G F(t) (a-a^{+}),
\end{eqnarray}
\begin{equation}
[a,a^+ ]= 1/N,\quad [J_{+},J_{-}]= 2 J_z /N,
\quad [J_{z},J_{\pm} ]=\pm J_{\pm} /N,
\end{equation}
where $a^{+}$, $a$ are the normalized operators of creation and
annihilation of photons in the cavity; $ J_{\alpha} =
N^{-1}\sum_{i=1}^{N}\sigma_{\alpha,i}\quad(\alpha=+,-,z)$
are collective atomic operators ($\sigma_{\alpha,i}$ are Pauli matrices
describing the $i$-th atom);
$\varrho=N/V$ is the density of TLS, $ d$ and $\omega_0$  are the dipole 
matrix element and transition frequency, respectively; 
$ G=c\omega_R /l\omega_c^2$, $ l$ is
characteristic sample size, $c$ is light velocity , $\omega_R =dE_0/\hbar$
is the
Rabi frequency. In (2)-(4) the case of exact resonance between atoms and
field is considered $\omega=\omega_0$. The atom-field coupling constant
in frequency units $\omega_c=(2\pi\varrho d^2 \omega_0/\hbar^2)^{1/2}$ is called
cooperative frequency [22] (in modern literature, $\omega_c$ is also called
``collective vacuum-field Rabi frequency'' [23]). It characterizes
the energy exchange from the atoms to  the field and back again. Such kind
of oscillations have been observed experimentally in both optical and
microwave domains [23-25].
\par
     One can easy see that eq. (3) represents the Hamiltonian of
the Jaynes-Cammings model [9] , and eq. (4) belongs to the type
``external source -- classical current'', according to the
classification suggested in [26].
\par
Different modifications of driven  Jaynes-Cammings model were introduced
in the
framework of investigation of the quantum field fluctuations including
squeezing [27]. But this activity was limited only by the integrable case
with regular dynamics. In contrast, the model (2)-(4) demonstrates both 
regular and chaotic dynamics.
\par
The quantum dynamics of our
system is completely described by  equation of motion for the density matrix
\begin{equation}
i\hbar\dot{\rho} = [H,\rho]
\end{equation}
     Now we introduce new operators
\begin{equation}
X_1=Re\,a = \frac{1}{2}(a^{+}+a),\quad  X_2=Im\,a=\frac{i}{2} (a^{+}-a)
\end{equation}
From the commutation relationships (5), we obtain the uncertainty
relation
\begin{equation}
\langle(\delta X_1)^2\rangle \langle(\delta X_2)^2\rangle
\geq\frac{1}{16 N}, \quad
 \delta X_j =X_j -\langle X_j\rangle
\end{equation}
The definition of squeezing has the form [16]
\begin{equation}
S=4 N \langle(\delta X_1)^2\rangle < 1
\end{equation}
\par			
     In order to study the dynamics of squeezing in the model
(2)-(4) at $N\gg 1$, it is natural to use the method of $ 1/N $ expansion.
It has been recently mentioned in [28] that the method can be
applied not only to the integrable but also to nonintegrable
systems. In accordance with this method [29], the natural states
for constructing the classical limit $ (N\rightarrow \infty)$ are generalized
coherent states. In our case, the state $\mid\psi\rangle$ is the product of 
the
Glauber coherent state $\mid\alpha\rangle$ [14] and of the spin coherent
state $\mid \beta\rangle$ [30-31]. For the model (2)-(4), the $1/N$ expansion is
constructed in a manner analogous to the expansion for the
Jaynes-Cummings model [18]. Therefore, I'd like to dwell here only
on some principal moments.
\par
     We define a generalized $P$-function as
\begin{equation}
\rho(t)=\int d^2 \alpha \, d^2 \beta \, P(\alpha,\beta,t)
\mid \psi\rangle\langle\psi\mid,
\end{equation}
$$
\int d^2 \alpha \equiv\int_{-\infty}^{+\infty} d(Re\,\alpha)
\int_{-\infty}^{+\infty} d(Im \,\alpha)
$$
Using the commutation relations between the Bose operators and
projectors [14] and the spin operators and projectors [31]
\begin{eqnarray}
 a^{+} \mid\alpha\rangle\langle\alpha\mid & = &
 U_{a^{+}}\mid\alpha\rangle\langle\alpha\mid,\nonumber \\
 a \mid\alpha\rangle\langle\alpha\mid & = &
 U_{a}\mid\alpha\rangle\langle\alpha\mid,\nonumber  \\
 \mid\alpha\rangle\langle\alpha\mid a^{+} & = &
 U_{a^{+}}^*\mid\alpha\rangle\langle\alpha\mid,  \\
 \mid\alpha\rangle\langle\alpha\mid a & = &
 U_{a}^*\mid\alpha\rangle\langle\alpha\mid, \nonumber
\end{eqnarray}
$$
U_{a^{+}}={\alpha}^* +\frac{1}{N} \frac{\partial}{\partial\alpha},\quad
U_{a}=\alpha
$$
\begin{eqnarray}
 J_{+} \mid\beta\rangle\langle\beta\mid &=&
 U_{J_{+}}\mid\beta\rangle\langle\beta\mid,\nonumber  \\
 J_{-}\mid\beta\rangle\langle\beta\mid &=&
 U_{J_{-}}\mid\beta\rangle\langle\beta\mid,\nonumber  \\
 \mid\beta\rangle\langle\beta\mid J_{+} &=&
 U_{J_{-}}^*\mid\beta\rangle\langle\beta\mid,  \\
  \mid\beta\rangle\langle\beta\mid J_{-} &=&
 U_{J_{+}}^*\mid\beta\rangle\langle\beta\mid,\nonumber
\end{eqnarray}
$$
U_{J_{+}}=\frac{\beta^*}{1+\mid\beta\mid^2}+
\frac{1}{N}\frac{\partial}{\partial\beta},\quad
U_{J_{-}}=\frac{\beta}{1+\mid\beta\mid^2}-
\frac{\beta^2}{N}\frac{\partial}{\partial\beta}
$$
from eq.(6), we have the equation of motion for the function $P$
\begin{equation}
\frac{\partial P}{\partial t}=\omega_c\left[-\frac{\partial}{\partial\alpha}
(V_{\alpha} P)-\frac{\partial}{\partial\beta}(V_{\beta} P)+
\frac{\partial^2}{\partial\alpha\partial\beta}(W_{\alpha\beta} P)
\right] + \mbox{c.c.}
\end{equation}
\begin{equation}
V_{\alpha}=\frac{\beta}{1+\mid\beta\mid^2}+G F(t),\quad
V_{\beta}=-(\alpha+\alpha^*\beta^2),\quad W_{\alpha\beta}=-\beta^2/N
\end{equation}
It follows from (13) and (14) that the variables $\alpha$ and $\beta$ can be
considered to be real. From eq. (13) one can obtain the following
equations of motion for averages [18]
\begin{eqnarray}
\frac{d}{d t}\langle\varphi\rangle  =
\omega_c \langle V_{\varphi}\rangle,\\
\frac{d}{d t}\langle\delta\varphi\delta\varphi\prime\rangle =
\omega_c \langle V_{\varphi}\delta\varphi\prime\rangle+
\omega_c \langle \delta\varphi V_{\varphi\prime}\rangle+
\omega_c \langle W_{\varphi\varphi\prime}\rangle,
\end{eqnarray}
where $\delta\varphi=\varphi-\langle\varphi\rangle$, and $\varphi$ is one of
the variables $\alpha$ or $\beta$. However,
the equations (15) and (16) are not closed. So, let us use the
Taylor's expansion of the function $V_{\varphi}$
\begin{equation}
V_{\varphi}=\left( V_{\varphi}\right)_{\langle\varphi\rangle}+
\sum_{\varphi\prime} \left(\frac{\partial V_{\varphi}}{\partial\varphi\prime}
\right)_{\langle\varphi\rangle} \delta\varphi\prime+\cdots
\end{equation}
and the analogous equation for $W_{\varphi\varphi\prime}$. Due to the 
uncertainty
relation of the type (8), it is clear that $\delta\varphi(t=0)\simeq N^{-1/2}$.
Therefore, substituting eq.(17) into (15) and (16) one can obtain
a closed system of equations for observed values in any order
over $1/N$. In a zero order, the equation of motion takes the form
\begin{equation}
\frac{d}{d t}\langle\varphi\rangle=
\omega_c \left( V_{\varphi}\right)_{\langle\varphi\rangle}+O(1/N)
\end{equation}
These equations are equivalent to the equations obtained from a
coupled Maxwell-Bloch system in [12]. After replacing the
variables $\beta=\tan(x/2)$ [30], eq. (18) can be written as follows
\begin{equation}
\ddot{x}+\omega_c^2 \sin x =-2 G\omega_c^2 F(t),
\end{equation}
where
\begin{equation}
\alpha=p/2,\quad p=-\frac{1}{\omega_c}\dot{x},\quad
\langle J_{+}\rangle=-\frac{1}{2}\sin x,\quad \langle J_z\rangle=
-\frac{1}{2}\cos x
\end{equation}
It is known already [1,2,12] that eq. (19) assumes both regular and
chaotic behaviour. The conditions for the transition to chaos in
(18),(19) will be discussed in the next Section.
\par
     In the first order in $1/N$, one can obtain from eq. (16) with an
account of (17)
\begin{eqnarray}
\frac{d}{d t}\langle(\Delta p)^2\rangle & = &
2\omega_c\cos x \langle\Delta p\Delta x\rangle,\nonumber \\
\frac{d}{d t}\langle(\Delta x)^2\rangle &=&-
2\omega_c\langle\Delta p\Delta x\rangle,\\
\frac{d}{d t}\langle\Delta p\Delta x\rangle &=&
\omega_c\cos x \langle(\Delta x)^2\rangle-
\omega_c\langle(\Delta p)^2\rangle,\nonumber
\end{eqnarray}
where
\begin{equation}
\langle(\Delta p)^2\rangle=4\langle(\delta\alpha)^2\rangle+2/N,\quad
\langle(\Delta x)^2\rangle=4 R^2\langle(\delta\beta)^2\rangle+2/N,
\end{equation}
$$
\langle\Delta p\Delta x\rangle
=4 R\langle\delta\alpha\delta\beta\rangle,\quad
R=(1+\beta^2)^{-1}
$$
In the present paper, we shall focuse our attention at the conditions when
initially field is in the coherent state and the atoms are in the ground
state $(J_z=-1/2) $ corresponding to $x(0)=0 $ and $p(0)=p_0$.
The condition for squeezing in variables (22) takes the form
\begin{equation}
S=N \langle(\Delta p)^2\rangle < 3
\end{equation}	
\par	
     Before we start investigating the nonlinear dynamics of the
systems (19) and (21), let us discuss the condition for validity
of the semislassical approach to squeezing.
\par
     The time-scale for validity of semiclassical approach and
$1/N$ expansion in cooperative optical system were studied systematically in
[32]. For regular dynamics, this time-scale has a power dependence on $N$.
In contrast, it is proportional to $\log N$ for a chaotic dynamics [32].
\par
     Here we only estimate a required number of TLS, when the
semiclassical description (18),(19) and (21) can be used to
describe the dynamics of squeezing. Introduce the ``convergence radius''
\begin{equation}
d(t)=\left[ \langle(\delta\alpha)^2\rangle+
\langle(\delta\beta)^2\rangle \right]^{1/2}\simeq
\left[ \langle(\Delta p)^2\rangle+\langle(\Delta x)^2\rangle\right]^{1/2}
\ll 1,
\end{equation}
where $d(0)\simeq N^{-1/2}$. If $d(t)\ll 1$, then $1/N$ expansion and 
consequently equations (18),(19) and (21) are
correct. The behaviour of $d(t)$ differs considerably
for regular and chaotic dynamics. Equations (21), in fact,
coincide\footnote[3]{\noindent The difference is in the fact that in
the definition of
Lyapunov exponent only the linearization near $x(t)$, $p(t)$ is
considered and not the behaviour of the linear fluctuations as in
(21). However, in this case the difference is insignificant. }
  with the equations arising in the definition of the
maximal Lyapunov exponent [1]. It is known [1,2] that for
the regular dynamics $d(t)$ increase according to a power law and
for the chaotic motion it is exponential
\begin{equation}
d(t)=d(0) \exp(\lambda\omega_c t),
\end{equation}
where $\lambda>0$ is the maximal Lyapunov exponent. The dependence (25)
is related to the presence of the strong local instability of the
chaotic motion. We will consider the effective squeezing to occur
for the time $t_1$, being of the order of several $\omega_c^{-1}$,
and $1/N$
expansion is correct, if $d\lesssim 0.01$ . With this assumption,
from (25)
it follows that one need to have $N\simeq 10^5\div 10^6$  for correct
semiclassical description of squeezing at chaos conditions.
This simple estimation is in a good agreement with the results of numerical
simulation [20].
\section{Chaos}
     In this section, following [12], we will briefly discuss the
conditions for transition to chaos in the model (18), (19). We
will need this information below.
\par
     The motion of a pendulum without perturbation $(G=0)$ is
periodic, and in phase plane it has two types of the fixed
points: elliptic, with coordinates $p=0$, $x=2\pi n$ $(n=0,\pm 1,\pm 2,
\ldots)$
corresponding (see (20)) to the initial populations of de-excited
levels of TLS, and hyperbolic, with coordinates $p=0$, $x=\pi (2n+1)$
$(n=0,\pm 1,\ldots)$ corresponding to the complete filling of the upper
levels of TLS. The pendulum separatrix (a special trajectory in
the phase plane, separating the vibrational and rotational motion
and going through the hyperbolic points) corresponds to the
complete energy transformation from the atoms to the field and
back again.
\par
     The dynamics of (18),(19) at $G\neq 0$ depends considerably on the
number of harmonics in the spectrum of $F(t)$. Let $F(t)=\sin \Omega t$. The
criteria for the transition to chaos obtained by means of the
Chirikov's resonance overlapping method [1-4] are different in
two limiting cases: $\kappa\gg 1$ and $\kappa\ll 1$,
$\kappa\equiv 2 G\omega_c^2 /\Omega^2$ .\\
1) If $\kappa\ll 1$ and for $\Omega > \omega_c$, a narrow stochastic layer
is found in the
vicinity of the separatrix; the rest of the phase space is filled
by periodic trajectories. At $\Omega\lesssim\omega_c$
in the vicinity of the
separatrix there appears a broad stochastic layer which occupies
the major part of the phase space, except the area in the
vicinity of the elliptic point.\\
2) If $ \kappa\gg 1$, the criterion for the transition to chaos
\begin{equation}
K>1,\quad K=\mbox{const}\frac{\omega_c}{\Omega\kappa^{1/4}},\quad
\mbox{const}\simeq 10
\end{equation}
can be easily satisfied. The oscillation amplitude in this case
may be rather large
\begin{equation}
\mid p_{\rm max}\mid\simeq\frac{\kappa\Omega}{\omega_c}=
\frac{2 G\omega_c}{\Omega}
\end{equation}
According to eq. (26) at $\Omega\rightarrow 0$ chaos is always present,
but in this
case the diffusion rate falls in proportional to $\Omega$, and the
chaotic motion in the system is slow (adiabatic chaos). Of great
importance is the fact that in the case under consideration the
transition to chaos is possible, if only the lower levels of TLS
are initially occupied.
\par
     A numerical analysis shows that at $\kappa\simeq 1$ the dynamics of the
system is qualitatively analogous to the case of $\kappa\gg 1$.
\par
     When the number of harmonics for $F(t)$ is increased, the
chaotic properties of the system become strongly.
\par
     Fig.~1~(a,b) and 2 show examples of chaotic, regular oscillations
and adiabatic chaos, respectively $(\tau=\omega_c^{-1} t)$. Fig 3 demonstrates
the presence of the local instability in the chaotic motion (a)
and its absence in the regular motion (b).
\par
     Dissipative analog of such model was considered in [19]
(both atomic and field dumping were included). In that case chaos
is transient.
\section{Squeezing and Chaos}
\par
    Let us consider now the dynamics of light squeezing. As the
system (18),(19) is Hamiltonian, the Liouville theorem is always
true. However, the change in the shape of phase volume for
regular and chaotic dynamics is different. It is known [1,2] that
for chaos the change of the phase volume shape is the strongest
and fastest. This is related to the fact that in the Hamiltonian
systems the mechanisms for the onset of chaos are the phase
contour stretching in one direction (and squeezing in another)
and its further folding. This procedure of squeezing and folding
is repeated several times. Eventually, the structure of the phase
volume contour becomes complicated, and its perimeter length
increases exponentially with time. The time intervals, when full
squeezing is possible, become very short and alternate
irregularly due to the chaotic motion. The dynamics of phase
volume contour for chaos is very sensitive to any changes in the
initial conditions and the system's parameters. This peculiarity
of the chaotic system is naturally manifested in the behaviour of
the system (21) determining the dispersion behaviour.
\par
     Though in the evolution process the phase volume shape of a
nonlinear system with regular dynamics also undergoes
considerable deformation, the stretching is less and rather
regular. The increase in length of the phase volume contour has a
power dependence. Therefore, the time intervals of squeezing have
to be long even for large times. The regular character of motion
has also to result in a weak dependence of the intervals on the
parameters and initial conditions.
\par
     These simple qualitative considerations have been verified
numerically. The squeezing (23) has been determined by solving the
system (19) and (21) when
$F(t)=\sin \Omega t$. The Runge-Kutta method of the fourth order has been
applied with an accuracy up to $10^{-9}\div 10^{-13}$. The precision of these
calculations has been monitored by checking that the
time-invariant of motion
\begin{equation}
 L=p/2-\cos x +2 x G\sin\psi+\Omega I,
\end{equation}
$$
\frac{d x}{d\tau}=\frac{\partial L}{\partial p},\quad
\frac{d p}{d\tau}=-\frac{\partial L}{\partial x},\quad
\frac{d\psi}{d\tau}=\Omega,\quad
\frac{d I}{d\tau}=-\frac{\partial L}{\partial\psi},\quad
I(0)=0,\quad \tau=\omega_c t
$$
is satisfied. Fig~4a shows the largest squeezing $S_{\rm min}$ for the
time $\tau=10\omega_c^{-1}$  as a function of the external perturbation $G$, and
Fig~5a shows the same as a function of the external field
frequency $\Omega$. The increase of $d$ during the same time
characterizes a degree of the dynamical instability (Figs. 4b and
5b). The conclusion about the character of oscillations in the
system (regular (R), chaotic (C), adiabatic chaos (AC)) have been
made on the basis of a phase portrait and the behaviour of $d(t) $
for times $\tau=200$. A boundary in the space of parameters between
chaos and adiabatic chaos is, naturally, quite relative. It can
be easily seen from figures that the degree of squeezing is
larger for chaos. The value of squeezing in this case makes up
$10^{-2}\div 10^{-3}$.
\par
     Several comparatively low values of squeezing $(S_{\rm min}\simeq 0.1)$
observed under chaos for some parameter values have been due to
the following reasons:\\
1) Weak statistical properties of the chaotic oscillations ( e.g.,
point 1 in figures 4 and 5) which are, as rule, related to the
fact that these parameters are in the vicinity of the area of
parameters corresponding to the regular dynamics.\\
2) Adiabatic chaotic dynamics. A characteristic period of the
oscillations in this case (see Fig. 2) is compared with the time
during which the squeezing was determined $(\tau_1 =10)$.\\
3) Anomalously strong instability of motion for $\tau\leq 10$ (see point 2
in figure 4a,b). The analysis of the behaviour $S(t)$ at this
parameter value shows that after a short fall $S(t)$ quickly
increases during the interval $(\tau_1 =10)$. This is due to fast and
strong deformation of the phase volume. However, strong
squeezing would be also possible further, e.g. for $\tau_1 =20$
 $S_{\rm min}\simeq 2.6\cdot 10^{-3}$.
\par
     Fig.6 shows temporary intervals of squeezing $(S<3)$, when the
dynamics of the system is regular or chaotic. It is seen that the
intervals are quite long and regular even at $\tau>10$ for the regular
dynamics (a) and short and irregular for the chaotic behaviour
(d). It is noteworthy that at $\tau\gtrsim 30$ the intervals of squeezing
become very short and their finding in numerical calculation
makes difficulties. The time interval of squeezing for the
chaotic motion is more sensitive to slight changes of the initial
conditions than for the regular one (compare (b) and (e), (c) and
(f)).
\par
     Thus, the chaotic dynamics leads to the short, irregular and
unstable squeezing at $\tau>10$. It should be noted that for $N\gg 1$ the
value of the maximal squeezing does not, in fact depend on $N$.
\section{Discussion}
     Thus, the main result of the paper is somewhat paradoxical:
a nonlinear system, which is in fact a generator of noise, may be
more effective in suppressing quantum noise. The maximal
squeezing for the chaotic motion may be by a factor of $10^3$  more
than for the regular dynamics.
\par
     We believe that the effect discussed may be observed in a
modification of the experiments on squeezing, when the light
interacts with an ensemble of two-level atoms [33]. The
amplitude modulation of the external field leading to chaos may
be not only sinusoidal but may also represent a sequence of short
light pulses. Such a sequence could be generated by a mode locked
laser.
\par
     It should be also noted that in this paper a ring cavity is
considered. But it can be shown that the transition to chaos is
also possible for other cavity configurations. For example, in
the case of a Fabry-Perot cavity eq. (19) has to be replaced by
the pendulum type equation with a Bessel function instead of
sinus (see the second reference in [12]). Such an equation has also
chaotic solutions, but Hamiltonian chaos is transient [12].
\par
     Besides our model, an analogous scheme can be applied to
other models of cooperative optical system with chaotic dynamics
in the semiclassical limit:\\
1) The BZT model [8] and also its generalization for the
multi-level systems [34]. In these models the global chaos is
possible if only the RWA is violated.\\
2) The Shepelyansky model [10]: an ensemble of 3-level systems
interacting with two modes of self-consistent field. The global
chaos is possible within the framework of the RWA when the dipole
moments of one- and two photon transitions are commensurable.\\
3) The ensemble of TLS interacting with the self-consistent field
and the resonance external field of constant amplitude [11,13]. The
transition to chaos is also possible within the framework of the
RWA. According to classification of [26] this model belongs to
the type ``external source -- classical field''.\\
Consequently, in all the cases one should expect an increase in
squeezing when the conditions for chaos are satisfied.
\par
     Another class of models assuming the $1/N$ expansion and
chaotic dynamics are those models which describe the parametric interaction
of light waves. This can be easily understood if to take into
account the fact that any cooperative optic process may
correspond to a parametric process (e.g., see [35]).
At present time,  the several situations are known when multiple parametric
interaction of both scalar light waves [36] and waves of different
polarizations [37] appears at chaotic spatial evolution.
\par
The consideration of squeezing at chaotic light waves evolution will be
the subject of our future publication.
\section*{Acknowledgements}
I am grateful to G. P. Berman for useful discussion and
attraction of my attention to ref. 32. I also thank A. R. Kolovsky
for fruitful discussion. This work was partially supported by the
Russian Fund for Basic Research (grant 94-02-04410),
EC-Russia grant INTAS (94-2058) and Krasnoyarsk Regional Science Fund.
\section*{References}
\begin{description}
\item[1.]  M. Liberman and A. Lichtenberg, {\it Regular and Stochastic
            Motion}, (Springer-Verlag, New York, 1983).
\item[2.] G. M. Zaslavsky, {\it Chaos in Dynamic Systems}, (Harwood, New
           York, 1985).
\item[3.] B. V. Chirikov, F. M. Izrailev and D. L. Shepelyansky, Soviet
           Scientific Reviews C, 2 (1981) 209.
\item[4.] G. M. Zaslavsky, Phys. Rep. 80 (1981) 175.
\item[5.] {\it Chaos and Quantum Physics}, ed. by M. J. Giannoni, A. Voros
            and J. Zinn-Justin, Les Houches Session $LIL$ 1989,
	    (Elsevier, Amsterdam, 1991).
\item[6.] T. Dittrich and R. Graham, Ann. of Phys., 200 (1990) 363.
\item[7.] P. W. Milonni, M.-L. Shih and J. R. Ackerhalt 1987, {\it Chaos in
           Laser-Matter Interactions}, (World Scientific, Singapoore, 1987);
           J. R. Ackerhalt, P. W. Milonni and M.-L. Shih, Phys. Rep.,
	   128 (1985) 205.
\item[8.] P. I. Belobrov, G. M. Zaslavsky and G. Kh. Tartakovsky, Sov. Phys.
           JETP, 44 (1976) 945.
\item[9.] E.T. Jaynes and F. W. Cummings, Proc. IEEE, 51 (1963) 89.
\item[10.] D. L. Shepelyansky, Phys. Rev. Lett., 57 (1986) 1815.
\item[11.] K. N. Alekseev and G. P. Berman, Sov. Phys. JETP, 65 (1987) 1115.
\item[12.] K. N. Alekseev and G. P. Berman, Sov. Phys. JETP, 67 (1988) 1762;
            78 (1994) 296.
\item[13.] D. D. Holm, G. Kova\u{c}i\u{c} and B. Sundaram, Phys. Lett. A,
            154 (1991) 346; D. D. Holm and G. Kova\u{c}i\u{c}, Physica D,
	    56 (1992) 270.	
\item[14.] J. Perina, {\it Quantum Statistics of Linear and Nonlinear
            Optical Phenomena}, (Reidel, Dordrecht, 1984).
\item[15.] D. F. Walls, Nature, 280 (1979) 451.
\item[16.] D. F. Smirnov and A. S. Troshin, Sov. Phys. Usp., 30 (1987) 851.
\item[17.] S. Reynaud, A. Heidmann, E. Giacobino and C. Fabre,  Quantum
            Fluctuation in Optical System, in {\it Progress in Optics XXX},
	    ed. E. Wolf, (Elsevier, Amsterdam, 1992).
\item[18.] A. Heidmann, J. M. Raimond, S. Reynaud and N. Zagury, Opt.
            Commun., 54 (1985) 189.
\item[19.] V. G. Benza and S. W. Koch, Phys. Rev. A, 35 (1987) 174.
\item[20.] K. N. Alekseev, Kirensky Institute Preprint No. 674F, 1991
            (unpublished); K. N. Alekseev, G. P. Berman and D. D. Holm,
	    {\it Hamiltonian Optical Chaos: Classical, Semiclassical
	    and Quantum (review)}, 1994, to be publ.
\item[21.] R. Bonifacio and L. A. Lugiato, Phys. Rev. A, 18 (1978) 1129;
            M. B. Spencer and W. E. Lamb (Jr), Phys. Rev. A, 5 (1972) 884.
\item[22.] A. P. Kazantsev and V. S. Smirnov, Sov. Phys. JETP, 19 (1964) 130;
            A. I. Alekseev, Yu. A. Vdovin and V. M. Galitsky,
	    {\it Ibid.},19 (1964) 220;
            F.T. Arecchi, V. Degiorgio and C. G. Someda, Phys. Lett.,
	    27 (1968) 588.
\item[23.] D. Meschede, Phys. Rep., 211 (1992) 201, sec. 6.2
\item[24.] H. J. Kimble, Structure and Dynamics in Cavity Quantum
            Electrodynamics, in {\it Advanced in Atomic, Molecular and
	    Optical Physics, Supplement 2}, (Academic Press, Amsterdam, 1994).
\item[25.] V. V. Vasil'ev, V. S. Egorov, A. N. Feodorov and
            I. A. Chekhonin, Optica i Spectr., 76 (1994) 146 [in Russian].	
\item[26.] R. Gilmore and C. M. Bowden, J. Math. Phys., 17 (1976) 1617.
\item[27.] H. J. Carmichael, Phys. Rev. Lett., 55 (1985) 2790;
            P. Alsing, D.-S. Guo and H. J. Carmichael, Phys. Rev. A,
	    45 (1992) 5135; S. M. Dutra, P.L. Knight and H. Moya-Cessa,
	    {\it Ibid}, 49 (1994) 1993.
\item[28.] R. F. Fox and J. C. Eidson, Phys. Rev. A, 36 (1987) 4321.
\item[29.] L. G. Yaffe, Rev. Mod. Phys., 54 (1982) 407;
            Phys. Today, August issue,(1983) 50.
\item[30.] J. M. Radcliffe, J. Phys. A, 4 (1971) 313;
            F. T. Arecchi, E. Courtens, R. Gilmore and H. Thomas,
            Phys. Rev. A, 6 (1972) 2211.
\item[31.] L. M. Narducci, C. M. Bowden, V. Blumel, G. P. Garrazana and
            R. A. Tuft, Phys. Rev. A, 11 (1975) 973;
            J. R.  Glauber and F. Haake, Phys. Rev. A, 13 (1976) 357.
\item[32.] G. P. Berman, E. N. Bulgakov and G. M. Zaslavsky, Chaos,
            2 (1992) 257; G. P. Berman, E. N. Bulgakov and D. D. Holm,
	    Phys. Rev. A., 49 (1994) 4945; {\it Crossover-Time in
	    Quantum Boson and Spin Systems}, Lecture Notes in Physics,
	    (Springer-Verlag, Berlin, 1994).	
\item[33.] M. G. Raizen, L. A. Orozco, M. Xiao, T. L. Boyd and H. J. Kimble
            ,Phys. Rev. Lett., 59 (1987) 198;
	    D. M. Hope, H. A. Bachor, P. J. Manson and D. E. McClelland,
	    Phys. Rev. A, 46 (1993) R1181.
\item[34.] P. I. Belobrov, G. P. Berman, G. M. Zaslavsky and A. P.
            Slivinsky, Sov. Phys. JETP, 49 (1979) 993.
\item[35.] S. Stenholm, Phys. Rep., 6 (1973) 1.
\item[36.] K. N. Alekseev, G. P. Berman, A. V. Butenko, A. K. Popov,
            A. V. Shalaev and V. Z. Yakhnin, J. Mod. Opt., 37 (1990) 41;	
            N. V. Alekseeva, K. N. Alekseev, V. A. Balueva,G. P. Berman,
	    A. K. Popov and V. Z. Yakhnin, Opt. Quant. Electr.,
	    23 (1991) 603.
\item[37.] J. Yumoto and K. Otsuka, Phys. Rev. Lett., 54 (1985) 1806;
            S. Trillo, S. Wabnitz, Phys. Rev. A, 36 (1987) 3881.
\end{description}
\newpage
\vspace{1.5cm}
\epsfxsize=8cm
\hspace{3cm}
\epsfbox{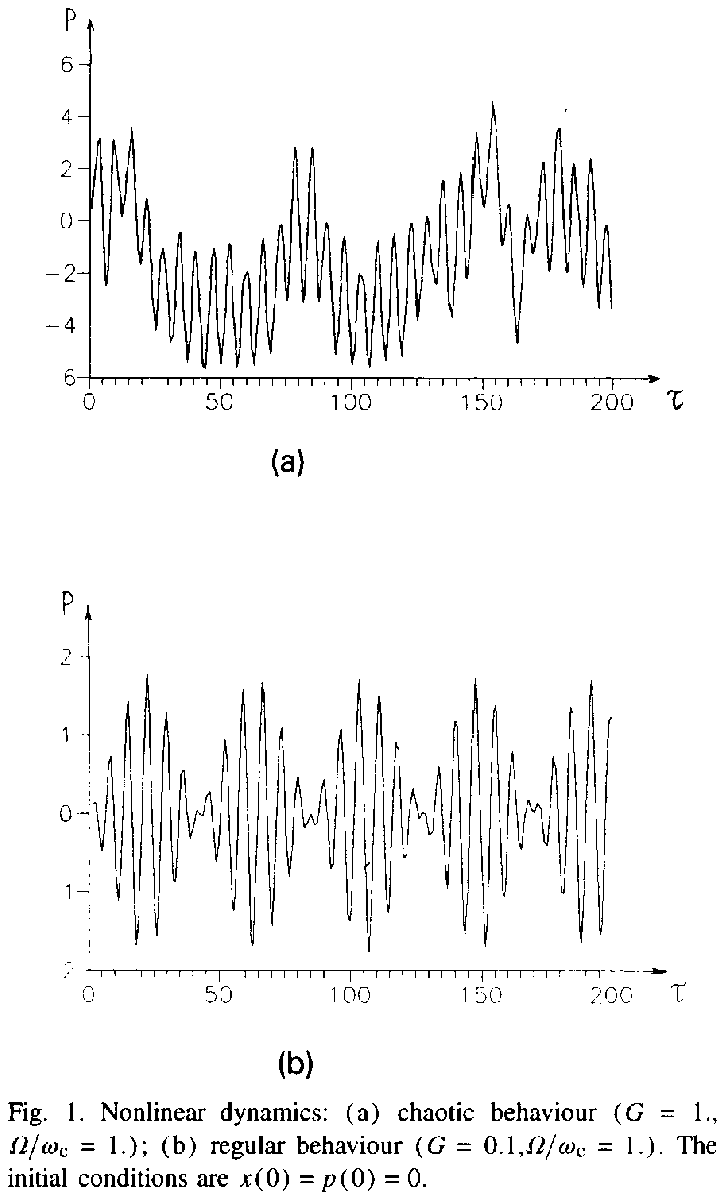}

\vspace{2cm}
\epsfxsize=9cm
\hspace{3cm}
\epsfbox{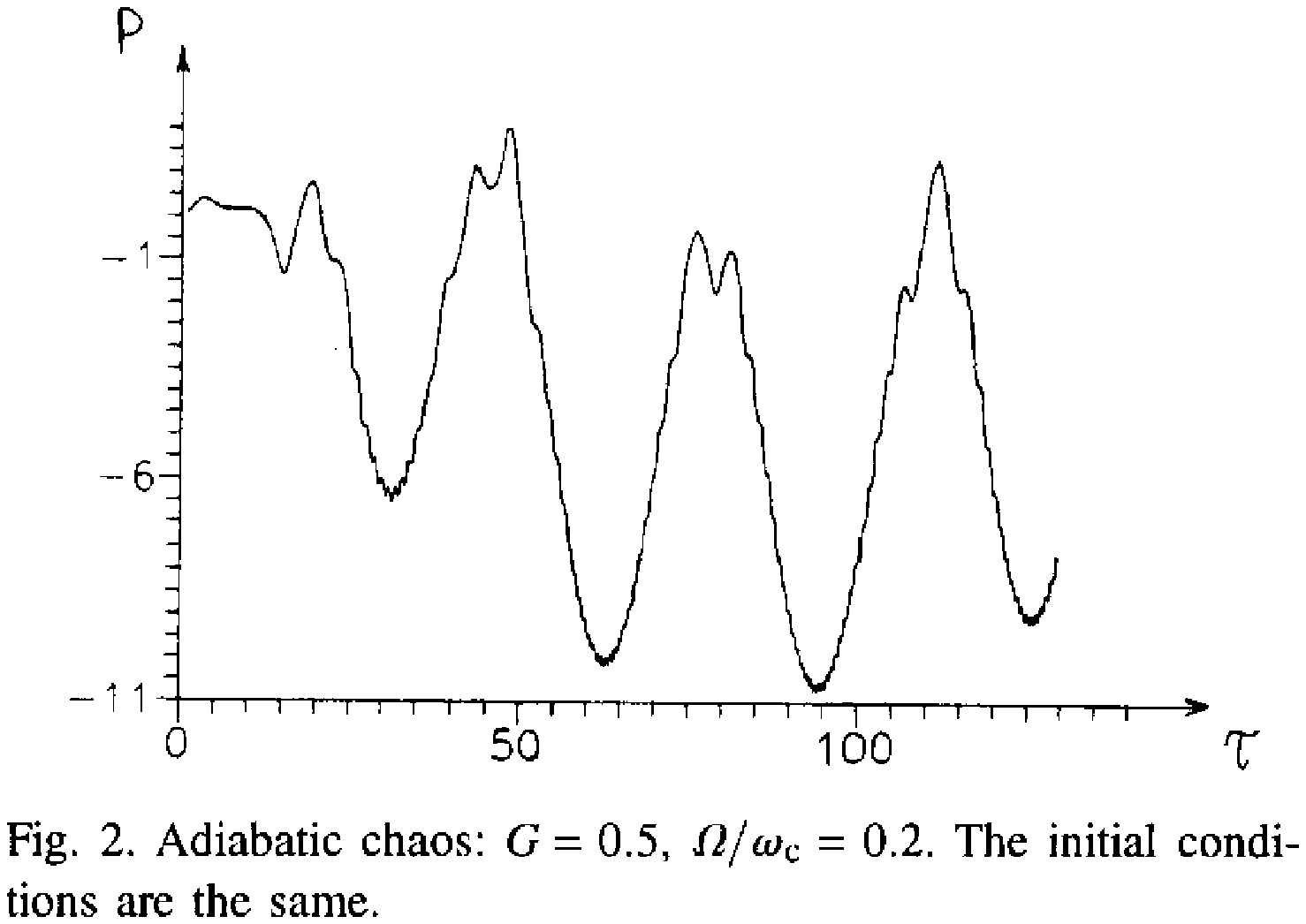}
\newpage
\vspace{2cm}
\epsfxsize=12cm
\hspace{1cm}
\epsfbox{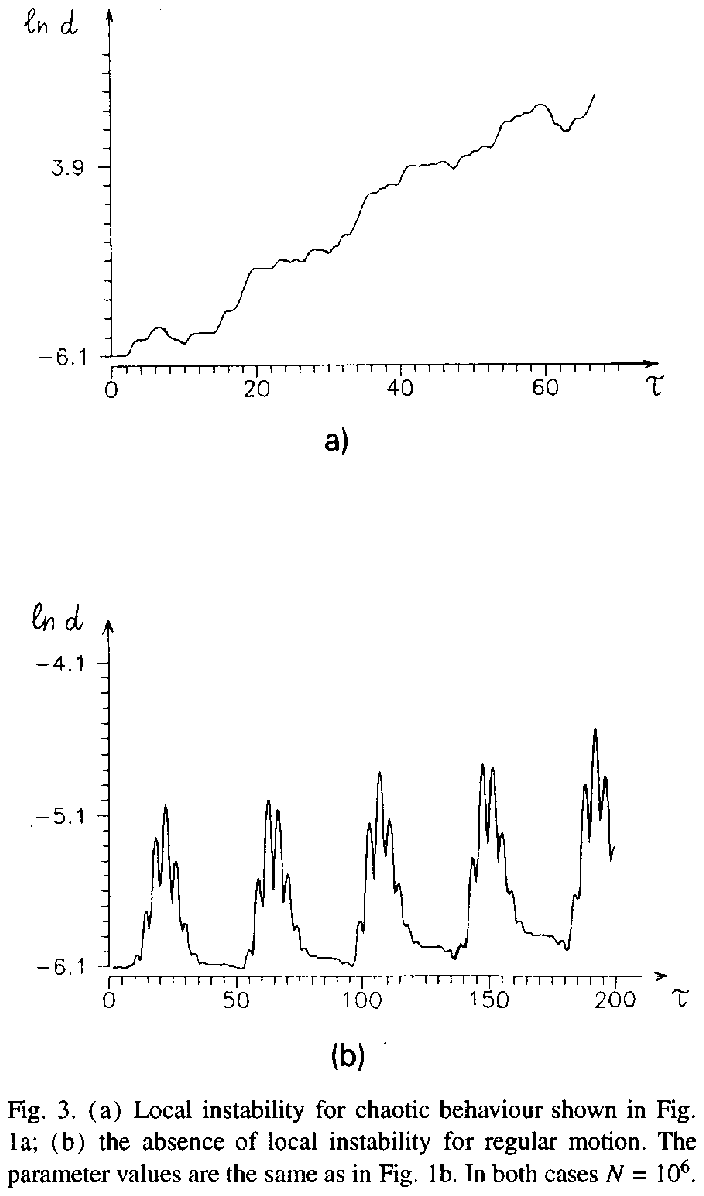}
\newpage

\vspace{2cm}
\epsfxsize=12cm
\hspace{1cm}
\epsfbox{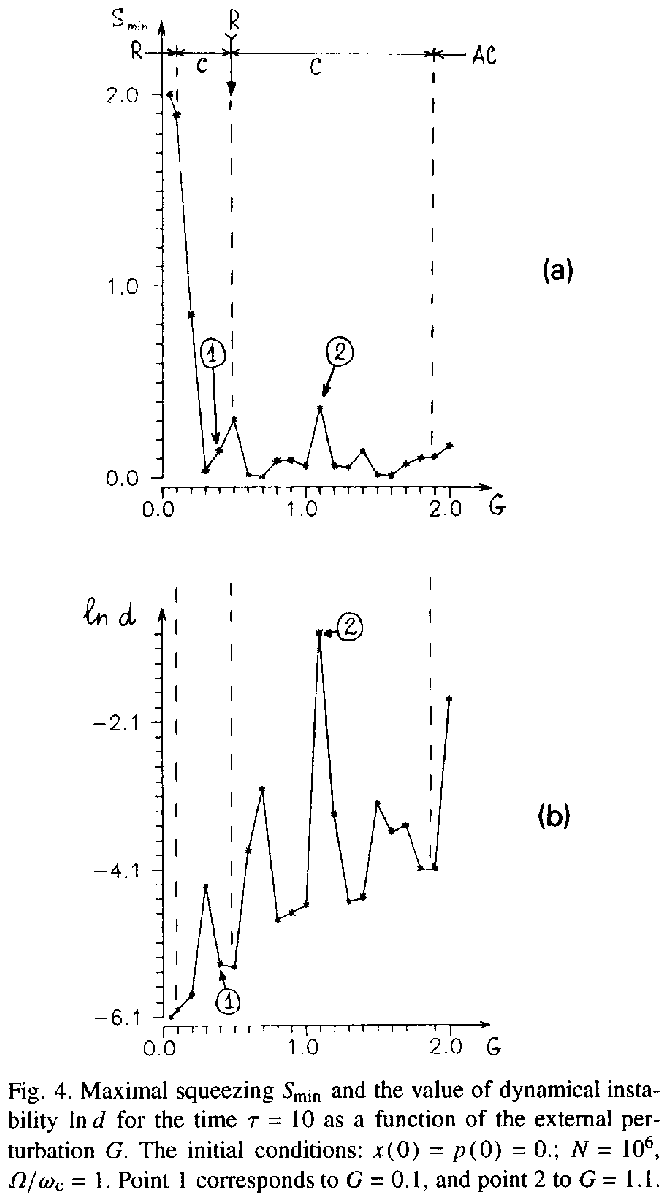}
\newpage
\vspace{1.5cm}
\epsfxsize=8cm
\hspace{4.5cm}
\epsfbox{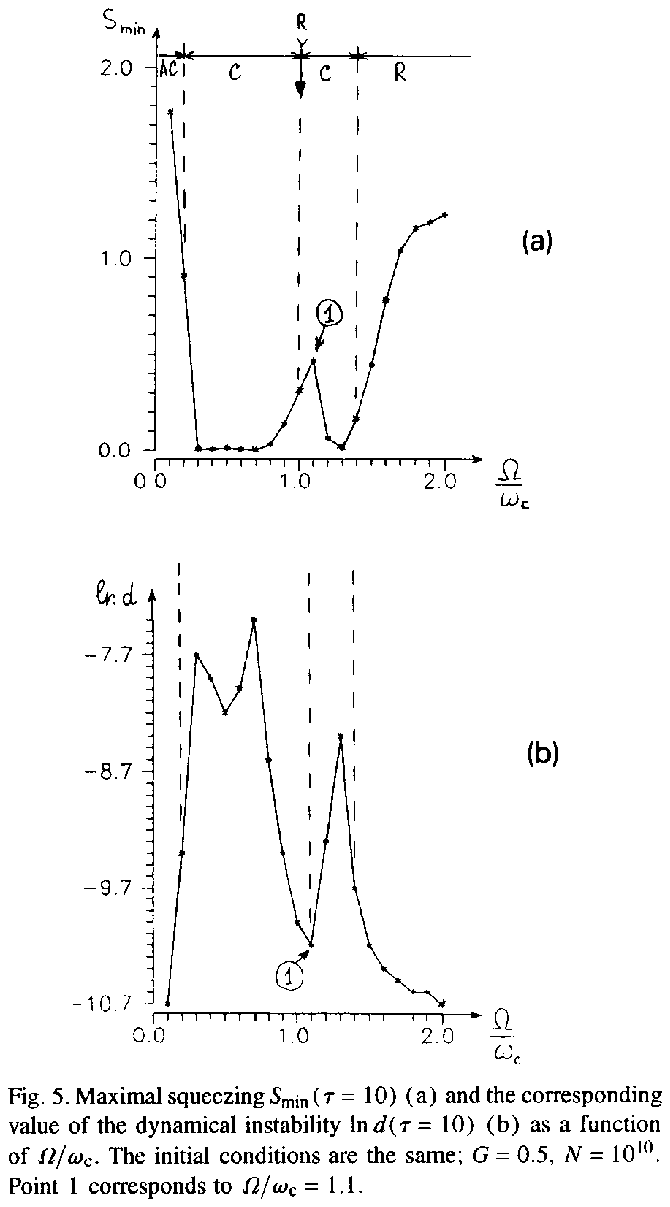}

\vspace{1cm}
\epsfxsize=17cm
\hspace{4cm}
\epsfbox{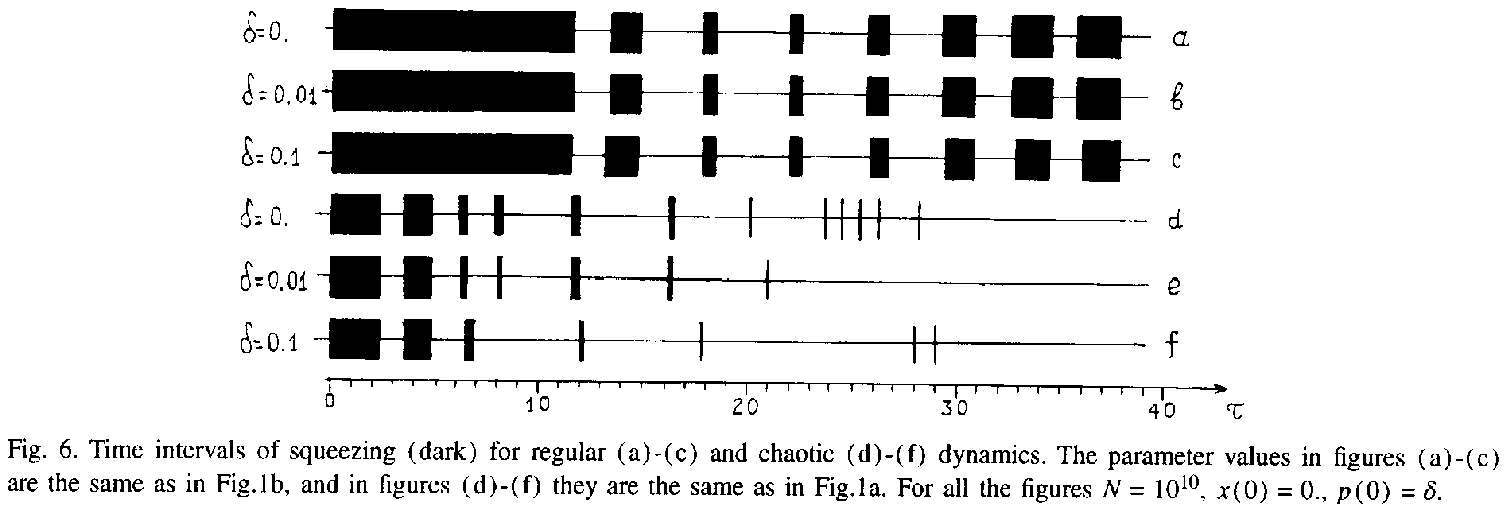}

\end{document}